\documentclass[doublespacing]{elsart}
\usepackage{graphicx}
\usepackage{amsmath}
\usepackage{amsfonts}
\usepackage{amssymb}

\begin{document}
\begin{frontmatter}
\title
{On a Separatrix in the Gravitational Collapse to an Overcritical Electromagnetic Black Hole}
\author[icra]{Remo Ruffini}\ead{ruffini@icra.it},
\author[icra]{Luca Vitagliano}\ead{vitagliano@icra.it},
\author[icra]{She-Sheng Xue}\ead{xue@icra.it}
\address
[icra]{ICRA and
Physics Department, University of Rome ``La Sapienza", 00185 Rome, Italy}
\begin{abstract}
The dynamical properties of an electron--positron--photon plasma created by the vacuum polarization process occurring around a charged gravitationally collapsing core of an initially neutral star are examined within the framework of General Relativity and Quantum Field Theory. The Reissner--Nordstr\"{o}m geometry is assumed to apply between the collapsing core and the oppositely charged remnant of the star. The appearance of a separatrix at radius $\bar{R}$, well outside the asymptotic approach to the horizon, is evidenced. The neutral electron--positron--photon plasma created at radii $r>\bar{R}$ self-propels outwards to infinity, following the classical PEM--pulse analysis \cite{RSWX99,RSWX00}. The plasma created at $r<\bar{R}$ remains trapped and follows the gravitational collapse of the core only contributing to the reduction of the electromagnetic energy of the black hole and to the increase of its irreducible mass. This phenomenon has consequences for the observational properties of gamma--ray bursts and is especially relevant for the theoretical prediction of the temporal and spectral structure of the short bursts.
\end{abstract}
\begin{keyword}
EMBH \sep electron--positron plasma \sep gravitational collapse \sep
gamma--ray bursts
\PACS95.30.Lz \sep97.60.Lf \sep98.70.Rz
\end{keyword}
\end{frontmatter}

The formulation of the physics of the \emph{dyadosphere} of an electromagnetic
black hole (EMBH) has been until now approached by assuming the vacuum
polarization process \emph{\`{a} l\`{a}} Sauter--Heisenberg--Euler--Schwinger
\cite{S31,HE35,S51} in the field of an already formed Kerr--Newmann
\cite{DR75} or Reissner--Nordstr\"{o}m black hole \cite{PRX98,RV02}. This
acausal approach is certainly valid in order to describe the overall
energetics and the time development of the gamma--ray bursts (GRBs) reaching a
remarkable agreement between the observations and the theoretical prediction,
in particular with respect to: a) the existence of a proper gamma--ray burst
(P--GRB) \cite{RBCFX01a}, b) the afterglow detailed luminosity function and
spectral properties \cite{RBCFX03a,RBCFVX03,RBCFX03b} and c) the relative
intensity of the P-GRB to the afterglow \cite{RBCFX01b,RBCFX03a,RBCFVX03}.

This acausal approach has to be improved by taking into account the causal
dynamical process of the formation of the dyadosphere as soon as the detailed
description on timescales of $10^{-4}-10^{-3}$s of the P--GRB are considered.
Such a description leads to theoretical predictions on the time variability of
the P--GRB spectra which may become soon testable by a new class of specially
conceived space missions.

It is the aim of this letter to report progress in this theoretically
challenging process which is marked by distinctive and precise quantum and
general relativistic effects. These new results have been made possible by the
recent progress in Refs.~\cite{CRV02}, \cite{RV02} and especially
\cite{RVX03}. There it was demonstrated the intrinsic stability of the
gravitational amplification of the electromagnetic field at the surface of a
charged star core collapsing to an EMBH. The $e^{+}e^{-}$ plasma generated by
the vacuum polarization process around the core is entangled in the
electromagnetic field \cite{RVX03b}. The $e^{+}e^{-}$ pairs do thermalize in
an electron--positron--photon plasma on a time scale $10^{2}-10^{4}$ times
larger than $\hbar/m_{e}c$ \cite{RVX03}, where $c$ is the speed of light and
$m_{e}$ the electron mass. As soon as the thermalization has occurred, a
dynamical phase of this electrically neutral plasma starts following the
considerations already discussed in \cite{RSWX99,RSWX00}. While the temporal
evolution of the $e^{+}e^{-}\gamma$ plasma takes place, the gravitationally
collapsing core moves inwards, giving rise to a further amplified
supercritical field, which in turn generates a larger amount of $e^{+}e^{-}$
pairs leading to a yet higher temperature in the newly formed $e^{+}%
e^{-}\gamma$ plasma. We report, in the following, progress in the
understanding of this crucial dynamical process: the main difference from the
previous treatments is the fact that we do not consider an already formed EMBH
but we follow the dynamical phase of the formation of dyadosphere and of the
asymptotic approach to the horizon by examining the time varying process at
the surface of the gravitationally collapsing core.

The space--time external to the surface of the spherically symmetric
collapsing core is described by the Reissner-Nordstr\"{o}m geometry \cite{P74}
with line element
\begin{equation}
ds^{2}=-\alpha^{2}dt^{2}+\alpha^{-2}dr^{2}+r^{2}d\Omega^{2},\label{ds}%
\end{equation}
with $d\Omega^{2}=d\theta^{2}+\sin^{2}\theta d\phi^{2}$, $\alpha^{2}%
=\alpha^{2}\left(  r\right)  =1-2M/r+Q^{2}/r^{2}$, where $M$ and $Q$ are the
total energy and charge of the core as measured at infinity. On the core
surface, which at the time $t_{0}$ has radial coordinate $r_{0}$, the
electromagnetic field strength is $\mathcal{E}=\mathcal{E}\left(
r_{0}\right)  =Q/r_{0}^{2}$. The equation of core's collapse is (see
\cite{CRV02}):
\begin{equation}
\tfrac{dr_{0}}{dt_{0}}=-\tfrac{\alpha^{2}\left(  r_{0}\right)  }{H\left(
r_{0}\right)  }\sqrt{H^{2}\left(  r_{0}\right)  -\alpha^{2}\left(
r_{0}\right)  }\label{Motion}%
\end{equation}
where $H\left(  r_{0}\right)  =\tfrac{M}{M_{0}}-\tfrac{M_{0}^{2}+Q^{2}}%
{2M_{0}r_{0}}$ and $M_{0}$ is the core rest mass. Analytic expressions for the
solution of Eq.~(\ref{Motion}) were given in \cite{CRV02}. We here recall that
the dyadosphere radius is defined by $\mathcal{E}\left(  r_{\mathrm{ds}%
}\right)  =\mathcal{E}_{\mathrm{c}}=$ $m_{e}^{2}c^{3}/e\hbar$ \cite{PRX98} as
$r_{\mathrm{ds}}=\sqrt{eQ\hbar/m_{e}^{2}c^{3}}$, where $e$ is the electron
charge. In the following we assume that the dyadosphere starts to be formed at
the instant $t_{\mathrm{ds}}=t_{0}\left(  r_{\mathrm{ds}}\right)  =0$.

Having formulated the core collapse in General Relativity in Eq.~(\ref{Motion}%
), in order to describe the quantum phenomena, we consider, at each value of
$r_{0}$ and $t_{0}$, a slab of constant coordinate thickness $\Delta r$ small
in comparison with $r_{\mathrm{ds}}$ and larger than $\hbar/m_{e}c^{2}$. All
the results will be shown to be independent on the choice of the value of
$\Delta r$. In each slab the process of vacuum polarization leading to
$e^{+}e^{-}$ pair creation is considered. As shown in \cite{RVX03,RVX03b} the
pairs created oscillate \cite{KESCM91,KESCM92,CEKMS93,BMP...99} with
ultrarelativistic velocities and partially annihilate into photons; the
electric field oscillates around zero and the amplitude of such oscillations
decreases with a characteristic time of the order of $10^{2}-10^{4}$
$\hbar/m_{e}c^{2}$. The electric field is effectively screened to the critical
value $\mathcal{E}_{\mathrm{c}}$ and the pairs thermalize to an $e^{+}%
e^{-}\gamma$ plasma. While the average of the electric field $\mathcal{E}$
over one oscillation is $0$, the average of $\mathcal{E}^{2}$ is of the order
of $\mathcal{E}_{c}^{2}$, therefore the energy density in the pairs and
photons, as a function of $r_{0}$, is given by \cite{RV02}
\begin{equation}
\epsilon_{0}\left(  r_{0}\right)  =\tfrac{1}{8\pi}\left[  \mathcal{E}%
^{2}\left(  r_{0}\right)  -\mathcal{E}_{c}^{2}\right]  =\tfrac{\mathcal{E}%
_{c}^{2}}{8\pi}\left[  \left(  \tfrac{r_{\mathrm{ds}}}{r_{0}}\right)
^{4}-1\right]  .\label{eps0}%
\end{equation}
For the number densities of $e^{+}e^{-}$ pairs and photons at thermal
equilibrium we have $n_{e^{+}e^{-}}\simeq n_{\gamma}$; correspondingly the
equilibrium temperature $T_{0}$, which is clearly a function of $r_{0}$ and is
different for each slab, is such that
\begin{equation}
\epsilon\left(  T_{0}\right)  \equiv\epsilon_{\gamma}\left(  T_{0}\right)
+\epsilon_{e^{+}}\left(  T_{0}\right)  +\epsilon_{e^{-}}\left(  T_{0}\right)
=\epsilon_{0},\label{eq0}%
\end{equation}
with $\epsilon$ and $n$ given by Fermi (Bose) integrals (with zero chemical
potential):
\begin{align}
\epsilon_{e^{+}e^{-}}\left(  T_{0}\right)   &  =\tfrac{2}{\pi^{2}\hbar^{3}%
}\int_{m_{e}}^{\infty}\tfrac{\left(  E^{2}-m_{e}^{2}\right)  ^{1/2}}%
{\exp\left(  E/kT_{0}\right)  +1}E^{2}dE,\quad\epsilon_{\gamma}\left(
T_{0}\right)  =\tfrac{\pi^{2}}{15\hbar^{3}}\left(  kT_{0}\right)
^{4},\label{Integrals1}\\
n_{e^{+}e^{-}}\left(  T_{0}\right)   &  =\tfrac{1}{\pi^{2}\hbar^{3}}%
\int_{m_{e}}^{\infty}\tfrac{\left(  E^{2}-m_{e}^{2}\right)  ^{1/2}}%
{\exp\left(  E/kT_{0}\right)  +1}EdE,\quad n_{\gamma}\left(  T_{0}\right)
=\tfrac{2\zeta\left(  3\right)  }{\hbar^{3}}\left(  kT_{0}\right)
^{3},\label{Integrals2}%
\end{align}

where $k$ is the Boltzmann constantFrom the conditions set by Eqs.~(\ref{eq0}%
), (\ref{Integrals1}), (\ref{Integrals2}), we can now turn to the dynamical
evolution of the $e^{+}e^{-}\gamma$ plasma in each slab. We use the covariant
conservation of energy momentum and the rate equation for the number of pairs
in the Reissner--Nordstr\"{o}m geometry external to the star core:
\begin{equation}
\nabla_{a}T^{ab}=0,\quad\nabla_{a}\left(  n_{e^{+}e^{-}}u^{a}\right)
=\overline{\sigma v}\left[  n_{e^{+}e^{-}}^{2}\left(  T\right)  -n_{e^{+}%
e^{-}}^{2}\right]  ,\label{na}%
\end{equation}
where $T^{ab}=\left(  \epsilon+p\right)  u^{a}u^{b}+pg^{ab}$ is the
energy--momentum tensor of the plasma with proper energy density $\epsilon$
and proper pressure $p$, $u^{a}$ is the fluid $4-$velocity, $n_{e^{+}e^{-}}$
is the number of pairs, $n_{e^{+}e^{-}}\left(  T\right)  $ is the equilibrium
number of pairs and $\overline{\sigma v}$ is the mean of the product of the
$e^{+}e^{-}$ annihilation cross-section and the thermal velocity of pairs. We
follow closely the treatment which we developed for the consideration of a
plasma generated in the dyadosphere of an already formed EMBH
\cite{RSWX99,RSWX00}. It was shown in \cite{RSWX99,RSWX00} that the plasma
expands as a pair--electromagnetic pulse (PEM pulse) of constant thickness in
the laboratory frame. Since the expansion, hydrodynamical timescale is much
larger than the pair creation ($\hbar/m_{e}c^{2}$) and the thermalization
($10^{2}-10^{4}\hbar/m_{e}c^{2}$) time-scales, in each slab the plasma remains
at thermal equilibrium in the initial phase of the expansion and the right
hand side of the rate Eq.~(\ref{na}) is effectively $0$, see Fig.~24 (second
panel) of \cite{RBCFVX03} for details.

If we denote by $\xi^{a}$ the static Killing vector field normalized at unity
at spacial infinity and by $\left\{  \Sigma_{t}\right\}  _{t}$ the family of
space-like hypersurfaces orthogonal to $\xi^{a}$ ($t$ being the Killing time)
in the Reissner--Nordstr\"{o}m geometry, from Eqs.~(\ref{na}), the following
integral conservation laws can be derived (see for instance \cite{D79,S60})
\begin{equation}
\int_{\Sigma_{t}}\xi_{a}T^{ab}d\Sigma_{b}=E,\quad\int_{\Sigma_{t}}%
n_{e^{+}e^{-}}u^{b}d\Sigma_{b}=N_{e^{+}e^{-}},\label{Ne}%
\end{equation}
where $d\Sigma_{b}=\alpha^{-2}\xi_{b}r^{2}\sin\theta drd\theta d\phi$ is the
vector surface element, $E$ the total energy and $N_{e^{+}e^{-}}$ the total
number of pairs which remain constant in each slab. We then have
\begin{equation}
\left[  \left(  \epsilon+p\right)  \gamma^{2}-p\right]  r^{2}=\mathfrak
{E},\quad n_{e^{+}e^{-}}\gamma\alpha^{-1}r^{2}=\mathfrak{N}_{e^{+}e^{-}%
},\label{ne}%
\end{equation}
where $\mathfrak{E}$ and $\mathfrak{N}_{e^{+}e^{-}}$ are constants and
\begin{equation}
\gamma\equiv\alpha^{-1}u^{a}\xi_{a}=\left[  1-\alpha^{-4}\left(  \tfrac
{dr}{dt}\right)  ^{2}\right]  ^{-1/2}%
\end{equation}
is the Lorentz $\gamma$ factor of the slab as measured by static observers. We
can rewrite Eqs.~(\ref{Ne}) for each slab as
\begin{align}
\left(  \tfrac{dr}{dt}\right)  ^{2} &  =\alpha^{4}f_{r_{0}},\label{eq17}\\
\left(  \tfrac{r}{r_{0}}\right)  ^{2} &  =\left(  \tfrac{\epsilon+p}%
{\epsilon_{0}}\right)  \left(  \tfrac{n_{e^{+}e^{-}0}}{n_{e^{+}e^{-}}}\right)
^{2}\left(  \tfrac{\alpha}{\alpha_{0}}\right)  ^{2}-\tfrac{p}{\epsilon_{0}%
}\left(  \tfrac{r}{r_{0}}\right)  ^{4},\label{eq18}\\
f_{r_{0}} &  =1-\left(  \tfrac{n_{e^{+}e^{-}}}{n_{e^{+}e^{-}0}}\right)
^{2}\left(  \tfrac{\alpha_{0}}{\alpha}\right)  ^{2}\left(  \tfrac{r}{r_{0}%
}\right)  ^{4}\label{eq19}%
\end{align}
where pedex $_{0}$ refers to quantities evaluated at selected initial times
$t_{0}>0$, having assumed $r\left(  t_{0}\right)  =r_{0}$, $\left.
dr/dt\right|  _{t=t_{0}}=0$, $T\left(  t_{0}\right)  =T_{0}$.

Eq.~(\ref{eq17}) is only meaningful when $f_{r_{0}}\left(  r\right)  \geq0$.
From the structural analysis of such equation it is clearly identifiable a
critical radius $\bar{R}$ such that:

\begin{itemize}
\item  for any slab initially located at $r_{0}>\bar{R}$ we have $f_{r_{0}%
}\left(  r\right)  \geq0$ for any value of $r\geq r_{0}$ and $f_{r_{0}}\left(
r\right)  <0$ for $r\lesssim r_{0}$; therefore a slab initially located at a
radial coordinate $r_{0}>\bar{R}$ moves outwards,

\item  for any slab initially located at $r_{0}<\bar{R}$ we have $f_{r_{0}%
}\left(  r\right)  \geq0$ for any value of $r_{+}<r\leq r_{0}$ and $f_{r_{0}%
}\left(  r\right)  <0$ for $r\gtrsim r_{0}$; therefore a slab initially
located at a radial coordinate $r_{0}<\bar{R}$ moves inwards and is trapped by
the gravitational field of the collapsing core.
\end{itemize}

We define the surface $r=\bar{R}$, the \emph{dyadosphere trapping surface
}(DTS). The radius $\bar{R}$ of DTS is generally evaluated by the condition
$\left.  \tfrac{df_{\bar{R}}}{dr}\right|  _{r=\bar{R}}=0$.
$\bar{R}$ is so close to the horizon value $r_{+}$ that the initial
temperature $T_{0}$ satisfies $kT_{0}\gg m_{e}c^{2}$ and we can obtain for
$\bar{R}$ an analytical expression. Namely the ultrarelativistic approximation
of all Fermi integrals, Eqs.~(\ref{Integrals1}) and (\ref{Integrals2}), is
justified and we have $n_{e^{+}e^{-}}\left(  T\right)  \propto T^{3}$ and
therefore $f_{r_{0}}\simeq1-\left(  T/T_{0}\right)  ^{6}\left(  \alpha
_{0}/\alpha\right)  ^{2}\left(  r/r_{0}\right)  ^{4}$ ($r\leq\bar{R}$).
The defining equation of $\bar{R}$, together with (\ref{eq19}), then gives
\begin{equation}
\bar{R}=2M\left[  1+\left(  1-3Q^{2}/4M^{2}\right)  ^{1/2}\right]  >r_{+}.
\end{equation}

\begin{figure}[th]
\begin{center}
\includegraphics[width=10cm]{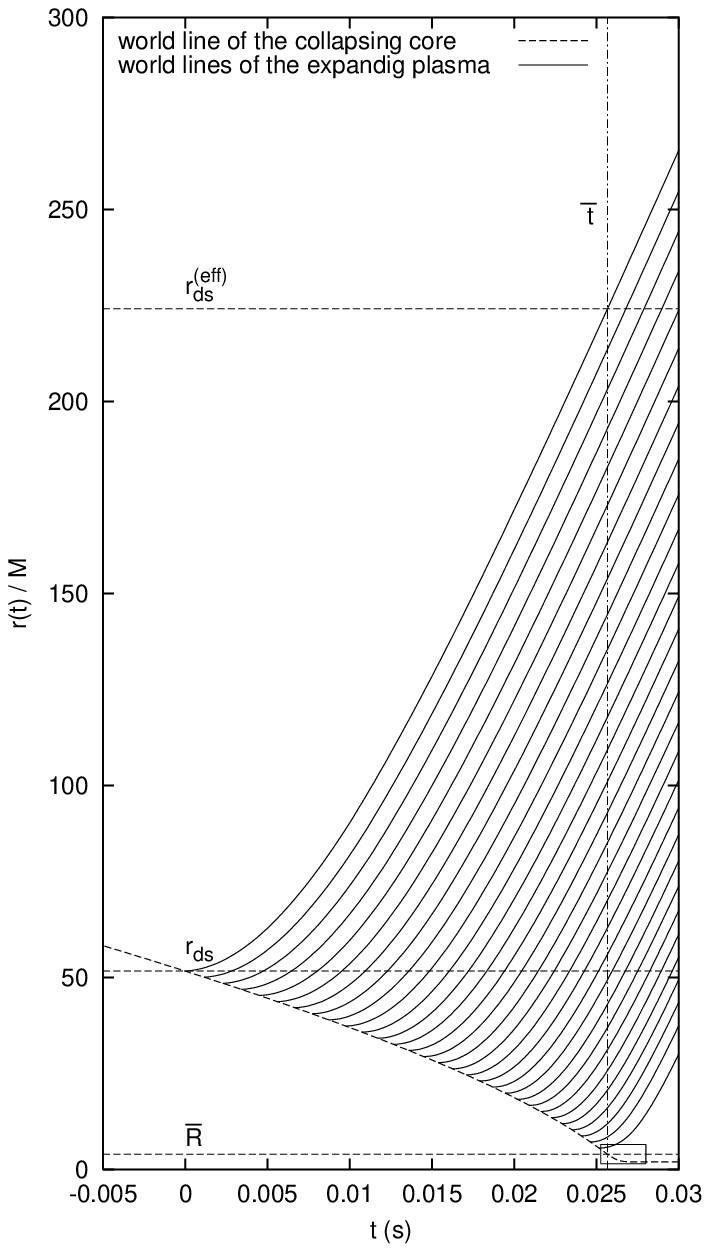}
\end{center}
\caption{World line of the collapsing charged core (dashed line) as derived
from Eq.~(\ref{Motion}) for an EMBH with $M=20M_{\odot}$, $Q=0.1M$; world
lines of slabs of plasma for selected radii $r_{0}$ in the interval $\left(
\bar{R},r_{\mathrm{ds}}\right)  $. At time $\bar{t}$ the expanding plasma
extends over a region which is almost one order of magnitude larger than the
dyadosphere. The small rectangle in the right bottom is enlarged in
Fig.~\ref{f2}.}%
\label{f1}%
\end{figure}

\begin{figure}[th]
\begin{center}
\includegraphics[width=13cm]{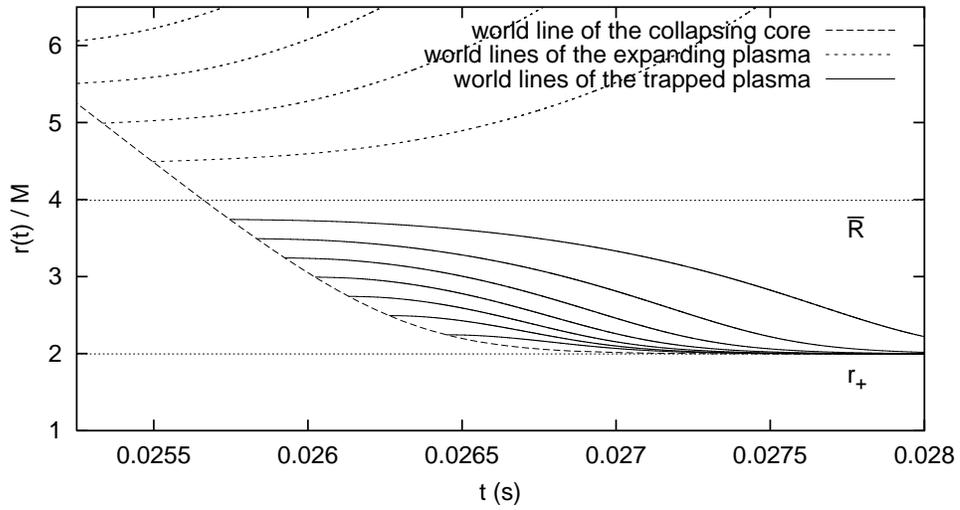}
\end{center}
\caption{Enlargement of the small rectangle in the right bottom of
Fig.~\ref{f1}. World--lines of slabs of plasma for selected radii $r_{0}$ in
the interval $\left(  r_{+,}\bar{R}\right)  $.}%
\label{f2}%
\end{figure}

In the case of an EMBH with $M=20M_{\odot}$, $Q=0.1M$, we compute:

\begin{itemize}
\item  the fraction of energy trapped in DTS:
\begin{equation}
\bar{E}=\int_{r_{+}<r<\bar{R}}\alpha\epsilon_{0}d\Sigma\simeq0.53\int
_{r_{+}<r<r_{\mathrm{ds}}}\alpha\epsilon_{0}d\Sigma;
\end{equation}

\item  the world--lines of slabs of plasma for selected $r_{0}$ in the
interval $\left(  \bar{R},r_{\mathrm{ds}}\right)  $ (see Fig.~\ref{f1});

\item  the world--lines of slabs of plasma for selected $r_{0}$ in the
interval $\left(  r_{+},\bar{R}\right)  $ (see Fig.~\ref{f2}).
\end{itemize}

\begin{figure}[th]
\begin{center}
\includegraphics[width=13cm]{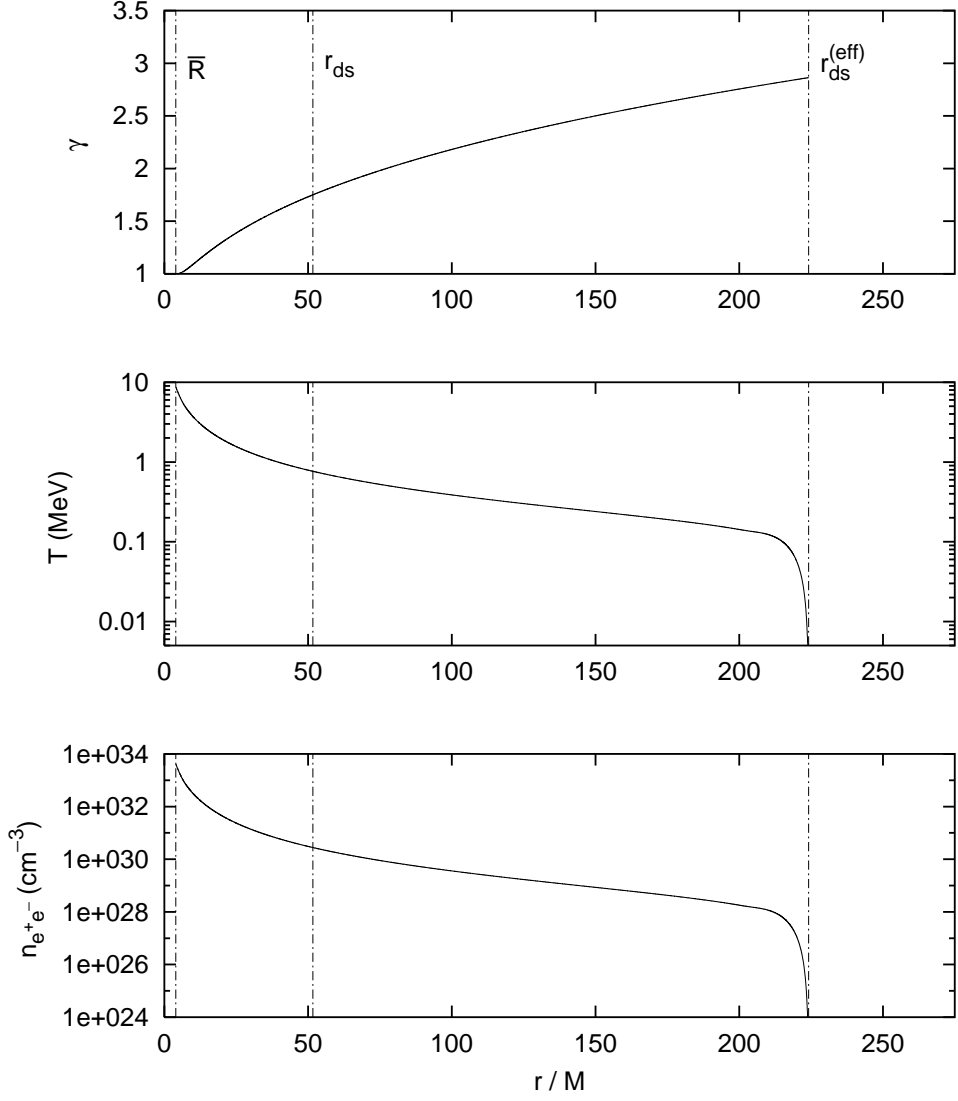}
\end{center}
\caption{Physical parameters in the effective dyadosphere: Lorentz $\gamma$
factor, proper temperature and proper $e^{+}e^{-}$ number density as functions
at time $\bar{t}$ for an EMBH with $M=20M_{\odot}$ and $Q=0.1M$.}%
\label{f3}%
\end{figure}

At time $\bar{t}\equiv t_{0}\left(  \bar{R}\right)  $ when the DTS is formed,
the plasma extends over a region of space which is almost one order of
magnitude larger than the dyadosphere and which we define as the
\emph{effective dyadosphere}. The values of the Lorentz $\gamma$ factor, the
temperature and $e^{+}e^{-}$ number density in the effective dyadosphere are
given in Fig.~\ref{f3}.

In conclusion we see how the causal description of the dyadosphere formation
can carry important messages on the time variability and spectral distribution
of the P--GRB due to quantum effects as well as precise signature of General Relativity.

\newpage

\newpage

\end{document}